\documentstyle{llncs}
\begin{document}
\title{QUANTUM CONVOLUTIONAL ERROR CORRECTION CODES}
\author{H. F. Chau\thanks{Electronic address: hfchau@hkusua.hku.hk}}
\institute{Department of Physics, University of Hong Kong, Pokfulam Road,
 Hong Kong}
\maketitle
\begin{abstract}
 I report two general methods to construct quantum convolutional codes for
 quantum registers with internal $N$ states. Using one of these methods, I
 construct a quantum convolutional code of rate 1/4 which is able to correct
 one general quantum error for every eight consecutive quantum registers.
\end{abstract}
\section{Introduction}
\label{S:Intro}
\par
 Quantum error correction code (QECC) is a succinct way to protect a quantum
 state from decoherence. The basic idea behind all QECC schemes is that by
 suitably encoding a quantum state in a larger Hilbert space $H$, and then
 later on measuring the wave function into certain subspace $C$ of $H$, it is
 possible to detect the kind of errors that have occurred. Finally, one can
 correct the error by applying a suitable unitary transformation to the
 orthogonal complement of $C$ according to the measurement result \cite{9-bit}.
 Many QECCs have been discovered in the last few years (see, for example,
 Refs.~\cite{AnE2,Cald3,Cald2,Cald1,9-reg,5-reg,Gott,5-bit,AnE1,N-add,9-bit,St1,St2,St3})
 and various theories on the QECC have also been developed (see, for example,
 Refs.~\cite{Cond3,AnE2,Cald2,9-reg,5-reg,Gott,Cond2,Cond1,AnE1,NonB,N-add,St2}).
 In particular, the necessary and sufficient condition for a QECC is
 \cite{Cond3,Cond2,Cond1}
\begin{equation}
 \langle i_{\rm encode} | {\cal A}^{\dag} {\cal B} | j_{\rm encode} \rangle =
 \Lambda_{{\cal A},{\cal B}} \,\delta_{ij} ~, \label{E:Condition}
\end{equation}
 where $|i_{\rm encode} \rangle$ denotes the encoded quantum state $|i\rangle$
 using the QECC; ${\cal A},{\cal B}$ are the possible errors the QECC can
 handle; and $\Lambda_{{\cal A},{\cal B}}$ is a complex constant independent of
 $|i_{\rm encode} \rangle$ and $|j_{\rm encode} \rangle$.
\par
 All QECCs discovered so far are block codes. That is, the original state ket
 is first divided into {\em finite} blocks of the same length. Each block is
 then encoded separately using a code which is {\em independent} of the state
 of the other blocks ({\em cf.} Refs.~\cite{Convolution2,Convolution1}).
 Besides block codes, convolutional codes are well known in classical error
 correction. Unlike a block code, the encoding operation depends on current as
 well as a number of past information bits \cite{Convolution2,Convolution1}.
 For instance, given a (possibly infinite) sequence of classical binary numbers
 $( a_1,a_2,\ldots ,a_m,\ldots )$, the encoding $( b_1,c_1,b_2,c_2,\ldots ,b_m,
 c_m,\ldots )$ with
\begin{equation}
 b_i = a_i + a_{i-2} \bmod 2, ~c_i = a_i + a_{i-1} + a_{i-2} \bmod 2
 \label{E:Classical_Conv}
\end{equation}
 for all $i$, and $a_{0} = a_{-1} = 0$ is an example of classical convolutional
 code that can correct up to one error for every four consecutive bits (see,
 for example, chap.~4 in Ref.~\cite{Convolution2} and
 Lemma~\ref{Lemma:21-encoder} in Section~\ref{S:QCC-CCC} for details).
\par
 In classical error correction, good convolutional codes often outperforms
 their corresponding block codes in the sense that they have higher encoding
 efficiencies \cite{Convolution2,Convolution1}. Thus, it is instructive to
 find quantum convolutional codes (QCC) and to analyze their performance. Here,
 I report two ways to construct QCCs. And from one of these methods, I
 construct a QCC of rate 1/4 that can correct one quantum error for every eight
 consecutive quantum registers (see Ref.~\cite{QCC} for more details).
\section{Constructing Quantum Convolutional Codes From Quantum Block Codes}
\label{S:QCC-QBC}
 In this Section, I report a general scheme to construct QCCs from quantum
 block codes (QBCs). But before doing so, let me first introduce some basic
 notations. Suppose each quantum register has $N$ orthogonal eigenstates, where
 $N$ is an integer greater than one. Then, the basis of a general quantum state
 making up of a collection of possibly infinite quantum registers can be chosen
 as $\{ | \vec{k} \rangle \} \equiv \{ |k_1,k_2,\ldots ,k_m,\ldots \rangle \}$,
 where $k_m \in \bbbz_N$ for all $m \in \bbbz$ with $N \geq 2$. Moreover, I
 abuse the notation by defining $k_m = 0$ for all $m \leq 0$. Finally, all
 additions and multiplications in all state kets below are modulo $N$.
\begin{definition}
 Let $|\vec{x}\rangle \equiv \sum_{k_1,k_2,\ldots} a_{k_1,k_2,\ldots} |k_1,k_2,
 \ldots,k_m,\ldots\rangle \equiv \sum_{\{ \vec{k} \}} a_{\vec{k}} |\vec{k}
 \rangle$ be a quantum state. Any quantum error can be regarded as an error
 operator ${\cal E}$ acting on this state. In particular, there is a {\bf spin
 flip error} occurring at quantum register $m$ (with respected to the basis $\{
 |\vec{k}\rangle \}$) if and only if ${\cal E} |\vec{x}\rangle = \sum_{\{
 \vec{k} \}} a_{\vec{k}} |k_1,k_2,\ldots ,k_{m-1},\tilde{k}_m,k_{m+1},\ldots
 \rangle$, where $\tilde{k}_m (k_m,{\cal E})$ is a $\bbbz_N$-function of $k_m$
 and ${\cal E}$. Moreover, a spin flip error is said to be {\bf additive}
 provided that $\tilde{k}_m (k_m,{\cal E}) = k_m + \alpha \bmod N$ for some
 $\alpha ({\cal E})$.
 \par
 Similarly, there is a {\bf phase shift error} occurring at quantum register
 $m$ (with respected to the basis $\{ |\vec{k}\rangle \}$) if and only if
 ${\cal E} |\vec{x}\rangle = \sum_{\{ \vec{k} \}} a_{\vec{k}} \,f(k_m,{\cal E})
 |\vec{k}\rangle$ for some complex-valued function $f(k_m,{\cal E})$ with
 $|f|^2 = 1$. Spin flip and phase shift errors occurring at more than one
 quantum register are defined in a similar way. \label{Def:Errors}
\end{definition}
\par
 With the above notations and definition in mind, a QBC and a QCC can be
 defined as follows:
\begin{definition}
 The linear map sending
 \begin{eqnarray}
  & & |\vec{k}\rangle \equiv |k_1,k_2,\ldots ,k_n\rangle \nonumber \\ &
  \longmapsto & \sum_{i_1,i_2,\ldots ,i_m} \!\!a^{(\vec{k})}_{i_1,i_2,\ldots ,
  i_m} |i_1,i_2,\ldots ,i_m\rangle \equiv \sum_{\{ \vec{i} \}}
  a^{(\vec{k})}_{\vec{i}} \,|\vec{i}\rangle \equiv |\vec{k}_{\rm encode}\rangle
  ~, \label{E:Encode_QBC}
 \end{eqnarray}
 where $a^{(\vec{k})}_{\vec{i}} \in \bbbc$, and $k_i\in \bbbz_N$ for all $i = 1
 ,2,\ldots ,N$ is said to be a {\bf quantum block code} (QBC) that can correct
 errors in the set $E$ if and only if Eq.~(\ref{E:Condition}) is satisfied for
 all ${\cal A}, {\cal B} \in E$. Since Eq.~(\ref{E:Encode_QBC}) encodes ever
 $n$ quantum registers to $m$ registers, the {\bf rate} of this code is,
 therefore, defined as $n/m$. In addition, one can encode the quantum state
 $\bigotimes_p | \vec{k}^{(p)} \rangle$ using the above QBC as $\bigotimes_p |
 \vec{k}^{(p)}_{\rm encode} \rangle$.
 \par
 On the other hand, if the encoding scheme expressed in
 Eq.~(\ref{E:Encode_QBC}) depends on current as well as past quantum states
 (that is, the coefficients $a^{(\vec{k})}_{\vec{i}}$ in
 Eq.~(\ref{E:Encode_QBC}) depend on more than one $\vec{k}^{(p)}$), then it is
 called a {\bf quantum convolutional code} (QCC). The rate of this
 convolutional code equals $n/m$ because it asymptotically encodes every $n$
 quantum registers as $m$ registers. \label{Def:QBC}
\end{definition}
\par
 With the above definitions in mind, one can construct a family of QCCs from a
 QBC as follows:
\begin{theorem}
 Given a QBC in Eq.~(\ref{E:Encode_QBC}) and a quantum state $|\vec{k}\rangle
 \equiv \bigotimes_{i=1}^{+\infty} |\vec{k}_i \rangle$ making up of possibly
 infinitely many quantum registers, then the mapping
 \begin{equation}
  |\vec{k}\rangle \equiv \bigotimes_{i=1}^{+\infty} |\vec{k}_i \rangle
  \longmapsto |\vec{k}_{\rm encode} \rangle \equiv \bigotimes_{i=1}^{+\infty}
  \left[ \,\sum_{\{ \vec{j}_i \}} a^{(\sum_p \mu_{ip} \vec{k}_p )}_{\vec{j}_i}
  \,|\vec{j}_i \rangle \right] ~, \label{E:QCC_From_QBC}
 \end{equation}
 forms a QCC of rate $n/m$ provided that the matrix $\mu_{ip}$ is invertible.
 This QCC handles errors in the set $E\otimes E\otimes E\otimes\cdots$.
 \label{Thrm:QCC_QBC}
\end{theorem}
\begin{proof}
 Let me consider the effects of errors ${\cal E} \equiv {\cal E}_1 \otimes
 {\cal E}_2 \otimes {\cal E}_3 \otimes\cdots$ and ${\cal E}' \equiv {\cal E}'_1
 \otimes {\cal E}'_2 \otimes {\cal E}'_3 \otimes\cdots$ in $E\otimes E\otimes
 E\otimes\cdots$ on the encoded quantum registers by computing $\langle
 \vec{k}'_{\rm encode} | {\cal E}'^{\dag} {\cal E} | \vec{k}_{\rm encode}
 \rangle$. From Eq.~(\ref{E:Condition}), I find that
 \begin{eqnarray}
  \langle \vec{k}'_{\rm encode} | {\cal E}'^{\dag} {\cal E} |
  \vec{k}_{\rm encode} \rangle & = & \prod_{i=1}^{+\infty} \left[ \,\sum_{\{
  \vec{j}_i, \vec{j}'_i \}} \bar{a}^{(\sum_{p'} \mu_{ip'} \vec{k}'_{p'}
  )}_{\vec{j}'_i} \,a^{(\sum_p \mu_{ip} \vec{k}_p )}_{\vec{j}_i} \,\langle
  \vec{j}'_i | {\cal E}'^{\dag}_i {\cal E}_i | \vec{j}_i \rangle \right]
  \nonumber \\ & = & \prod_{i=1}^{+\infty} \left[ \left\langle \left(
  {\textstyle \sum_p \mu_{ip} \vec{k}'_p} \right)_{\rm encode} \left|
  {\cal E}'^{\dag}_i {\cal E}_i \right| \left( {\textstyle \sum_p \mu_{ip}
  \vec{k}_p} \right)_{\rm encode} \right\rangle \right] \nonumber \\ & = &
  \prod_{i=1}^{+\infty} \left[ \,\delta_{\sum_p \mu_{ip} \vec{k}_p , \sum_p
  \mu_{ip} \vec{k}'_p} \,\Lambda_{{\cal E}_i, {\cal E}'_i} \right]
  \label{E:Condition_QBC_1}
 \end{eqnarray}
 for some constants $\Lambda_{{\cal E}_i,{\cal E}'_i}$ independent of $\vec{k}$
 and $\vec{k}'$. Because $\mu$ is invertible, it is clear that $\vec{k}_i =
 \vec{k}'_i$ for all $i\in\bbbz^+$ is the unique solution for the systems of
 linear equations $\sum_p \mu_{ip} \vec{k}_p = \sum_p \mu_{ip} \vec{k}'_p$.
 Consequently, $\langle \vec{k}'_{\rm encode} | {\cal E}'^{\dag} {\cal E} |
 \vec{k}_{\rm encode} \rangle = \delta_{\vec{k},\vec{k}'} \,\Lambda_{{\cal E},
 {\cal E}'}$, for some constant $\Lambda_{{\cal E},{\cal E}'}$ independent of
 $\vec{k}$ and $\vec{k}'$. Thus, the mapping in Eq.~(\ref{E:QCC_From_QBC}) is a
 QCC. \qed
\end{proof}
\par
 Now, let me uses Theorem~\ref{Thrm:QCC_QBC} to give an example of QCC.
\begin{example}
 Starting from the five qubit perfect code for $N = 2$
 \cite{Cald3,5-reg,5-bit}, Theorem~\ref{Thrm:QCC_QBC} implies that the
 following QCC can correct up to one error in every five consecutive qubits:
 \begin{eqnarray}
  |k_1,k_2,\ldots ,k_m,\ldots\rangle & \longmapsto & \bigotimes_{i=1}^{+\infty}
  \left[ \,\frac{1}{N^{3/2}} \!\sum_{p_i,q_i,r_i = 0}^{N-1} \!\!(-1)^{(k_i +
  k_{i-1})(p_i + q_i + r_i) + p_i r_i} \right. \nonumber \\ & & ~~~~\left.
  \raisebox{1.8em}{~} \!\!|p_i,q_i,p_i + r_i,q_i + r_i,p_i + q_i + k_i +
  k_{i-1} \rangle \right] \label{E:Perfect_QCC}
 \end{eqnarray}
 where $k_m \in \{ 0,1 \}$ for all $m\in\bbbz^+$. The rate of this code is
 1/5. \label{Ex:Perfect_QCC}
\end{example}
\par
 Although the QCC in Eq.~(\ref{E:Encode_QBC}) looks rather complicated, the
 actual encoding process can be performed readily. Since $\mu$ is invertible,
 one can reversibly map $|\vec{k}_1,\vec{k}_2,\ldots ,\vec{k}_n,\ldots\rangle$
 to $|\sum_p \mu_{1p} \vec{k}_p,\sum_p \mu_{2p} \vec{k}_p,\ldots ,\sum_p
 \mu_{np} \vec{k}_p,\ldots\rangle$ \cite{Pebble,SpaceTime,Reversible}. Then,
 one obtains the above five bit QCC by encoding each quantum register using
 various encoding procedures described in Refs.~\cite{Cond3,24P,5-reg,5-bit}.
\section{Constructing Quantum Convolutional Codes From Classical Convolutional
 Codes}
\label{S:QCC-CCC}
 In this Section, I report a general method to construct QCCs from classical
 convolutional codes. My construction is based on the following two technical
 lemmas which hold for both QBCs and QCCs:
\begin{lemma}
 Suppose the QECC
 \begin{equation}
  |\vec{k}\rangle \longmapsto \sum_{\{ \vec{j} \}}  a^{(\vec{k})}_{\vec{j}} \,|
  \vec{j}\rangle \label{E:Code_Flip}
 \end{equation}
 corrects (independent) additive spin flip errors in certain quantum registers.
 Then, the following QECC, which is obtained by discrete Fourier transforming
 every quantum register in Eq.~(\ref{E:Code_Flip}),
 \begin{equation}
  |\vec{k}\rangle \longmapsto \sum_{\{ \vec{j},\vec{p} \}}
  a^{(\vec{k})}_{\vec{j}} \prod_{i=1}^{+\infty} \left( \frac{1}{\sqrt{N}}
  \,\omega_N^{j_i p_i} \right) |\vec{p}\rangle \label{E:Code_Phase}
 \end{equation}
 corrects (independent) phase errors occurring in the same set of quantum
 registers. The converse is also true. \label{Lemma:Spin_Phase}
\end{lemma}
\begin{proof}
 Consider two arbitrary but fixed additive spin flip errors ${\cal E} \equiv
 \bigotimes_{i=1}^{+\infty} {\cal E}_i$ and ${\cal E}' \equiv
 \bigotimes_{i=1}^{+\infty} {\cal E}'_i$ acting on the code in
 Eq.~(\ref{E:Code_Flip}). I denote the set of all quantum registers affected by
 either one of the above spin flip errors and unaffected by both errors as $A$
 and $U$, respectively. Then Eqs.~(\ref{E:Condition})
 and~(\ref{E:Code_Flip}) imply that
 \begin{equation}
  \sum_{\{ \vec{j}, \vec{j}' \}} \left[ \,\bar{a}^{(\vec{k}')}_{\vec{j}'}
  \,a^{(\vec{k})}_{\vec{j}} \left( \prod_{i\in U} \delta_{j_i,j'_i} \right)
  \left( \prod_{i\in A} \langle j'_i | {\cal E}'^{\dag}_i {\cal E}_i | j_i
  \rangle \right) \right] = \delta_{\vec{k},\vec{k}'} \,\Lambda_{{\cal E},
  {\cal E}'} \label{E:Condition_Flip}
 \end{equation}
 for some constant $\Lambda_{{\cal E},{\cal E}'}$ independent of $\vec{k}$ and
 $\vec{k}'$.
 \par
 For additive spin errors, $\langle j'_i | {\cal E}'^{\dag}_i {\cal E}_i | j_i
 \rangle = \langle j'_i + \alpha'_i | j_i + \alpha_i \rangle = \delta_{j'_i +
 \alpha'_i , j_i + \alpha_i}$ for some constants $\alpha_i , \alpha'_i \in
 \bbbz_N$. In other words, $\langle j'_i | {\cal E}'^{\dag} {\cal E} | j_i
 \rangle$ is a binary function of $j_i - j'_i$ only. Thus,
 Eq.~(\ref{E:Condition_Flip}) still holds if I replace $\langle j'_i |
 {\cal E}'^{\dag}_i {\cal E}_i | j_i \rangle$ by a binary function $g (j_i -
 j'_i : i\in A)$. Moreover, the linearity of Eq.~(\ref{E:Condition_Flip})
 implies that the same equation holds if I replace $\langle j'_i |
 {\cal E}'^{\dag}_i {\cal E}_i | j_i \rangle$ by {\em any} complex-valued
 function $g$ taking arguments on $j_i - j'_i$ for all $i\in A$. That is to
 say,
 \begin{equation}
  \sum_{\{ \vec{j},\vec{j}' \}} \left[ \,\bar{a}^{(\vec{k}')}_{\vec{j}'}
  \,a^{(\vec{k})}_{\vec{j}} \left( \prod_{i\in U} \delta_{j_i,j'_i} \right)
  g(j_i - j'_i :i\in A) \right] = \delta_{\vec{k},\vec{k}'} \,\Lambda_g
  \label{E:Condition_Flip_General}
 \end{equation}
 for some complex-valued $\Lambda_g$ independent of $\vec{k}$ and $\vec{k}'$.
 Conversely, it is obvious that if $a^{(\vec{k})}_{\vec{j}}$ satisfies
 Eq.~(\ref{E:Condition_Flip_General}), then Eq.~(\ref{E:Code_Flip}) is a QECC
 that is capable of correcting additive spin flip errors. In other words,
 Eq.~(\ref{E:Condition_Flip_General}) is a necessary and sufficient condition
 for the QECC to correct additive spin flip errors.
 \par
 Now, I consider the actions of two phase shift errors ${\cal F}$ and
 ${\cal F}'$ acting on the {\em same} set of quantum registers as those in
 ${\cal E}$ and ${\cal E}'$, respectively. Then
 \begin{eqnarray}
  & & \langle \vec{k}'_{\rm encode} | {\cal F}'^{\dag} {\cal F} |
  \vec{k}_{\rm encode} \rangle \nonumber \\ & = & \sum_{\{ \vec{j},\vec{j}',
  \vec{p},\vec{p}' \}} \left[ \,\bar{a}^{(\vec{k}')}_{\vec{j}'}
  \,a^{(\vec{k})}_{\vec{j}} \,\omega_N^{\sum_i ( j_i p_i - j'_i p'_i )}
  \prod_{i\in U} \left( \frac{1}{\sqrt{N}} \,\delta_{p_i,p'_i} \right) \right.
  \nonumber \\ & & ~~~~\left. \times \prod_{i\in A} \left( \frac{1}{\sqrt{N}}
  \,\langle p'_i | {\cal F}'^{\dag}_i {\cal F}_i | p_i \rangle \right) \right]
  \nonumber \\ & = & \sum_{\{ \vec{j}, \vec{j}',\vec{p} \}} \left\{
  \,\bar{a}^{(\vec{k}')}_{\vec{j}'} \,a^{(\vec{k})}_{\vec{j}} \left( \prod_{i
  \in U} \delta_{j_i,j'_i} \right) \sum_{p'_i : i\in A} \!\left[
  \raisebox{1.8em}{~} \!\!\omega_N^{\sum_{i\in A} ( j'_i p'_i - j_i p_i ) }
  \right. \right. \nonumber    \\ & & ~~~~\left. \left. \times \prod_{i\in A}
  \left( \frac{1}{\sqrt{N}} \,\langle p'_i | {\cal F}'^{\dag}_i {\cal F}_i |
  p_i \rangle \right) \right] \right\} ~. \label{E:Condition_Phase_1}
 \end{eqnarray}
 For phase shift errors, $\langle p'_i | {\cal F}'^{\dag}_i {\cal F}_i | p_i
 \rangle = \delta_{p_i,p'_i} \,h$ for some complex-valued function $h$ of $p_i
 : i\in A$ with $|h|^2 = 1$. Consequently, Eq.~(\ref{E:Condition_Phase_1}) can
 be further simplified as
 \begin{eqnarray}
  & & \langle {\bf k}'_{\rm encode} | {\cal F}'^{\dag} {\cal F} |
  {\bf k}_{\rm encode} \rangle \nonumber \\ & = & \sum_{\{ \vec{j},\vec{j}',
  \vec{p} \}} \left[ \,\bar{a}^{(\vec{k}')}_{\vec{j}'}
  \,a^{(\vec{k})}_{\vec{j}} \left( \prod_{i\in U} \delta_{j_i,j'_i} \right)
  \,\omega_N^{\sum_{i\in A} p_i ( j'_i - j_i ) } \,h(p_i : i\in A) \right] ~,
  \label{E:Condition_Phase_2}
 \end{eqnarray}
 for some complex-valued function $h (p_i : i\in A)$. Summing over all the
 $p_i$s in Eq.~(\ref{E:Condition_Phase_2}), I obtain
 \begin{eqnarray}
  & & \langle {\bf k}'_{\rm encode} | {\cal F}'^{\dag} {\cal F} |
  {\bf k}_{\rm encode} \rangle \nonumber \\ & = & \sum_{\{ \vec{j},\vec{j}' \}}
  \left[ \,\bar{a}^{(\vec{k}')}_{\vec{j}'} \,a^{(\vec{k})}_{\vec{j}} \left(
  \prod_{i\in U} \delta_{j_i,j'_i} \right) \,h'( j_i - j'_i : i\in A ) \right]
  ~, \label{E:Condition_Phase_3}
 \end{eqnarray}
 for some complex-valued function $h' (j_i - j'_i : i\in A)$. Comparing
 Eqs.~(\ref{E:Condition_Flip_General}) and~(\ref{E:Condition_Phase_3}), one
 concludes that $\langle \vec{k}'_{\rm encode} | {\cal F}'^{\dag} {\cal F} |
 \vec{k}_{\rm encode} \rangle = \delta_{\vec{k},\vec{k}'} \,\Lambda_{{\cal F},
 {\cal F}'}$ for some $\Lambda_{{\cal F},{\cal F}'}$ independent of both
 $\vec{k}$ and $\vec{k}'$. Thus, the QECC given in Eq.~(\ref{E:Code_Phase})
 corrects the phase shift errors as promised.
 \par
 Conversely, from Eq.~(\ref{E:Condition_Phase_3}), one concludes that
 Eq.~(\ref{E:Code_Phase}) corrects phase errors if and only if
 \begin{equation}
  \sum_{\{ \vec{j},\vec{j}' \}} \left[ \,\bar{a}^{(\vec{k}')}_{\vec{j}'}
  \,a^{(\vec{k})}_{\vec{j}} \left( \prod_{i\in U} \delta_{j_i,j'_i} \right)
  \,h'( j_i - j'_i : i\in A ) \right] = \delta_{\vec{k},\vec{k}'}
  \,\Lambda_{h'} \label{E:Condition_Phase_General}
 \end{equation}
 for {\em any} complex-valued function $h' (j_i - j'_i : i\in A)$. Hence, from
 Eq.~(\ref{E:Condition_Flip_General}), one concludes that
 Eq.~(\ref{E:Code_Flip}) is able to correct additive spin flips errors. \qed
\end{proof}
\par
 In essence, Lemma~\ref{Lemma:Spin_Phase} tells us that the abilities to
 correct additive spin flip and phase shift form a dual pair under the discrete
 Fourier transform of quantum registers. An interesting case occurs when $N =
 2$. Here, additive spin flip is the only possible kind of spin flip error. As
 a result, the abilities to correct spin flip and phase shift errors in $N = 2$
 form a dual pair under Lemma~\ref{Lemma:Spin_Phase}. And this special form of
 Lemma~\ref{Lemma:Spin_Phase} was proven earlier by various authors (see, for
 example, Refs.~\cite{Cald3,Cald2,Cond2}).
\begin{corollary}
 If a QECC handles both spin flip and phase shift errors on the same set of
 quantum registers, then this QECC handles any general quantum errors occurring
 at the same set of quantum registers. \label{Cor:Phase_Spin}
\end{corollary}
\begin{proof}
 Combining Eqs.~(\ref{E:Condition_Flip_General})
 and~(\ref{E:Condition_Phase_2}), one knows that
 Eq.~(\ref{E:Condition_Flip_General}) holds for {\em any} complex-valued
 function $g(j_i,j'_i : i\in A)$. By putting $\langle j'_i | {\cal E}'^{\dag}_i
 {\cal E}_i | j_i \rangle = g(j_i,j'_i)$ for all $i\in A$, then one concludes
 that the above QECC is capable of correcting any general quantum errors as
 promised. \qed
\end{proof}
\begin{lemma}
 Suppose QECCs $C1$ and $C2$ handle phase shift and spin flip errors,
 respectively, for the same set of quantum registers. Then, pasting the two
 codes together by first encodes the quantum state using $C1$ then further
 encodes the resultant quantum state using $C2$, one obtains a QECC $C$ which
 corrects general errors in the same set of quantum registers.
 \label{Lemma:Paste}
\end{lemma}
\begin{proof}
 Clearly $C$ can handle spin flip errors occurring at the specified quantum
 registers. So from Corollary~\ref{Cor:Phase_Spin}, it remains to show that $C$
 corrects phase errors as well. Let the encodings for $C1$ and $C2$ be $|
 \vec{k}\rangle \longmapsto \sum_{\{ \vec{j} \}} a^{(\vec{k})}_{\vec{j}} |
 \vec{j}\rangle$ and $|\vec{j}\rangle \longmapsto \sum_{\{ \vec{p} \}}
 b^{(\vec{j})}_{\vec{p}} |\vec{p}\rangle$, respectively. Then using the same
 set of notations as in the proof of Lemma~\ref{Lemma:Spin_Phase}, one knows
 that
 \begin{eqnarray}
  & & \langle \vec{k}'_{\rm encode} | {\cal F}'^{\dag} {\cal F}|
  \vec{k}_{\rm encode} \rangle \nonumber \\ & = & \sum_{\{ \vec{j},\vec{j}',
  \vec{p},\vec{p}' \}} \left[ \,\bar{a}^{(\vec{k}')}_{\vec{j}'}
  \,a^{(\vec{k})}_{\vec{j}} \,\bar{b}^{(\vec{j}')}_{\vec{p}'}
  \,b^{(\vec{j})}_{\vec{p}} \left( \prod_{i\in U} \delta_{p_i,p'_i} \right)
  \left( \prod_{i\in A} \langle p'_i | {\cal F}'^{\dag}_i {\cal F}_i | p_i
  \rangle \right) \right] \nonumber \\ & = & \sum_{\{ \vec{j},\vec{j}',\vec{p},
  \vec{p}' \}} \left\{ \,\bar{a}^{(\vec{k}')}_{\vec{j}'}
  \,a^{(\vec{k})}_{\vec{j}} \,\bar{b}^{(\vec{j}')}_{\vec{p}'}
  \,b^{(\vec{j})}_{\vec{p}} \delta_{\vec{p},\vec{p}'} \left[ \prod_{i\in A}
  g_i(p_i : i\in A) \right] \right\} \nonumber \\ & = & \sum_{\{ \vec{j},
  \vec{j}',\vec{p} \}} \left\{ \,\bar{a}^{(\vec{k}')}_{\vec{j}'}
  \,a^{(\vec{k})}_{\vec{j}} \,\bar{b}^{(\vec{j}')}_{\vec{p}}
  \,b^{(\vec{j})}_{\vec{p}} \left[ \prod_{i\in A} g(p_i : i\in A) \right]
  \right\} \label{E:Phase_Com_1}
 \end{eqnarray}
 for some complex-valued functions $g_i (p_i : i\in A)$ for all $i\in A$.
 \par
 Since $C2$ handles spin flips, one demands that whenever $\vec{j} \neq
 \vec{j}'$,
 \begin{equation}
  \sum_{\{ \vec{p} \}} \,\bar{b}^{(\vec{j}')}_{\vec{p}}
  \,b^{(\vec{j})}_{\vec{p}} = 0 = \sum_{\{ \vec{p},\vec{p}' \}}
  \,\bar{b}^{(\vec{j}')}_{\vec{p}'} \,b^{(\vec{j})}_{\vec{p}} \,\langle
  \vec{p}' | {\cal E} | \vec{p} \rangle ~, \label{E:Sum_Spin_Flip_1}
 \end{equation}
 where ${\cal E}$ denotes a possible spin flip error that can be handled by the
 QECC $C2$. Consequently,
 \begin{equation}
  \sum_{\{ \vec{p} \}}' \,\bar{b}^{(\vec{j}')}_{\vec{p}}
  \,b^{(\vec{j})}_{\vec{p}} = 0 ~, \label{E:Sum_Spin_Flip_2}
 \end{equation}
 where the above primed sum is over either (1) all the $\vec{p}$ that is
 affected by the error ${\cal E}$, or (2) all the $\vec{p}$ that is unaffected
 by the error ${\cal E}$.
 \par
 From Eq.~(\ref{E:Sum_Spin_Flip_2}), it is easy to see that after summing over
 all $p_i$s in Eq.~(\ref{E:Phase_Com_1}), one will arrive at
 \begin{equation}
  \langle \vec{k}'_{\rm encode} | {\cal F}'^{\dag} {\cal F} |
  \vec{k}_{\rm encode} \rangle = \sum_{\{ \vec{j},\vec{j}' \}} \left[
  \,\bar{a}^{(\vec{k}')}_{\vec{j}'} \,a^{(\vec{k})}_{\vec{j}}
  \,\delta_{\vec{j},\vec{j}'} \left( \prod_{i\in A} h_i (j_i : i\in A) \right)
  \right] ~, \label{E:Phase_Com_2}
 \end{equation}
 for some complex-valued function $h_i (j_i : i\in A)$. As $C1$ handles phase
 shift, one concludes that $\langle \vec{k}'_{\rm encode} | {\cal F}'^{\dag}
 {\cal F} | \vec{k}_{\rm encode} \rangle = \delta_{\vec{k},\vec{k}'}
 \,\Lambda_{{\cal F},{\cal F}'}$. Hence, the Lemma is proved. \qed
\end{proof}
\par
 At this point, I would like to remark that the proof of the abilities to
 correct both spin flip and phase shift implies the ability to correct a
 general error for $N = 2$ can be found in Refs.~\cite{Cald3,Cald2,Ekert,St3}.
 Moreover, one should notice that the ordering of encoding in
 Lemma~\ref{Lemma:Paste} is important. Encoding first using a spin flip code
 followed by a phase shift code does {\em not}, in general, result in a general
 QECC. After proving the above two technical lemmas, I report a method to
 construct QECCs from classical codes.
\begin{theorem}
 Suppose $C$ is a classical (block or convolutional) code of rate $r$ that
 corrects $p$ (classical) errors for every $q$ consecutive registers. Then,
 $C$ can be extended to a QECC of rate $r^2$ that corrects at least $p$
 quantum errors for every $q^2$ consecutive quantum registers.
 \label{Thrm:Pasting}
\end{theorem}
\begin{proof}
 Suppose $C$ is a classical (block or convolutional) code. By mapping $m$ to
 $|m\rangle$ for all $m\in\bbbz_N$, $C$ can be converted to a quantum code for
 spin flip errors. Let $C'$ be the QECC obtained by Fourier transforming each
 quantum register of $C$. Then Lemma~\ref{Lemma:Spin_Phase} implies that $C'$
 is a code for phase shift errors. From Lemma~\ref{Lemma:Paste}, pasting codes
 $C$ and $C'$ together will create a QECC of rate $r^2$. Finally, the fact that
 $C'$ corrects at least $p$ quantum errors for every $q^2$ consecutive quantum
 registers follows directly from Corollary~\ref{Cor:Phase_Spin}. \qed
\end{proof}
\par
 Theorem~\ref{Thrm:Pasting} provides a powerful way to create high rate QECCs
 from high rate classical codes.
\begin{example}[(Shor)]
 Starting with the simplest classical majority block code of rate $1/3$,
 namely, $|k\rangle \longmapsto |k,k,k\rangle$ for $k=0,1$,
 Theorem~\ref{Thrm:Pasting} returns the famous Shor's single error correcting
 nine bit code \cite{9-bit} of rate $1/9$:
 \begin{equation}
  |k\rangle \longmapsto \sum_{p,q,r=0}^1 (-1)^{k(p+q+r)} \,|p,p,p,q,q,q,r,r,r
  \rangle ~. \label{E:Shor-9-bit}
 \end{equation}
\end{example}
\par
 Alternatively, one may start with a high rate classical convolutional code.
 One of the simplest codes of this kind is the $1/2$-rate code in
 Eq.~(\ref{E:Classical_Conv}). Being a non-systematic\footnote{That is, both
 $b_i$ and $c_i$ are not equal to $a_i$.} and non-catastrophic\footnote{That
 is, a finite number of channel errors does not create an infinite number of
 decoding errors.} code (see, for example, chap.~4 in Ref.~\cite{Convolution2}
 for details), it serves as an ideal starting point to construct good QCCs.
 First, let me write down this code in quantum mechanical form:
\begin{lemma}
 The rate $1/2$ QCC
 \begin{equation}
  \bigotimes_{i=1}^{+\infty} |k_i\rangle \longmapsto |\vec{k}_{\rm encode}
  \rangle \equiv \bigotimes_{i=1}^{+\infty} |k_i + k_{i-2},k_i + k_{i-1} +
  k_{i-2} \rangle ~, \label{E:21_QSpin}
 \end{equation}
 where $k_i \in\bbbz_N$ for all $i\in\bbbz^+$, can correct up to one spin flip
 error for every four consecutive quantum registers. \label{Lemma:21-encoder}
\end{lemma}
\begin{proof}
 Here, I give a ``quantum version'' of the proof. Using notations in the proof
 of Theorem~\ref{Thrm:QCC_QBC}, I consider $\langle \vec{k}'_{\rm encode} |
 {\cal E}'^{\dag} {\cal E} | \vec{k}_{\rm encode} \rangle$ again. Clearly, the
 worst case happens when errors ${\cal E}$ and ${\cal E}'$ occur at different
 quantum registers. And in this case, Eq.~(\ref{E:21_QSpin}) implies that
 exactly two of the following four equations hold:
 \begin{equation}
  \left\{ \begin{array}{rcl} k_{2i} + k_{2i-2} & = & k'_{2i} + k'_{2i-2} \\
  k_{2i} + k_{2i-1} + k_{2i-2} & = & k'_{2i} + k'_{2i-1} + k'_{2i-2} \\
  k_{2i+1} + k_{2i-1} & = & k'_{2i+1} + k'_{2i-1} \\ k_{2i+1} + k_{2i} +
  k_{2i-1} & = & k'_{2i+1} + k'_{2i} + k'_{2i-1} \end{array} \right.
  \label{E:Systems_1}
 \end{equation}
 for all $i\in\bbbz^+$. One may regard $k_i$s as unknowns and $k'_i$s as
 arbitrary but fixed constants. Then, by straight forward computation, one can
 show that picking {\em any} two equations out of Eq.~(\ref{E:Systems_1}) for
 each $i$ will form an invertible system with the unique solution $k_i = k'_i$
 for all $i\in\bbbz^+$. Therefore, $\langle \vec{k}'_{\rm encode} |
 {\cal E}'^{\dag} {\cal E} | \vec{k}_{\rm encode} \rangle = \delta_{\vec{k},
 \vec{k}'} \,\delta_{{\cal E},{\cal E}'}$ and hence this lemma is proved. \qed
\end{proof}
\begin{example}
 Theorem~\ref{Thrm:Pasting} and Lemma~\ref{Lemma:21-encoder} imply that
 \begin{eqnarray}
  \bigotimes_{i=1}^{+\infty} |k_i\rangle & \longmapsto &
  \bigotimes_{i=1}^{+\infty} \left[ \sum_{p_1,q_1,\ldots} \!\frac{1}{N} \,
  \omega_N^{(k_i + k_{i-2}) p_i + (k_i + k_{i-1} + k_{i-2}) q_i} \,|p_i +
  p_{i-1}, \right. \nonumber \\ & & ~~~~\left. p_i + p_{i-1} + q_{i-1},q_i +
  q_{i-1},q_i + q_{i-1} + p_i \rangle \raisebox{1.5em}{~} \!\!\right] ~,
  \label{E:14QCC}
 \end{eqnarray}
 where $k_i\in\bbbz_N$ for all $i\in\bbbz^+$, is a rate 1/4 QCC capable of
 correcting up to one quantum error for every sixteen consecutive quantum
 registers.
 \par
 In what follows, I show that this code can in fact correct up to one quantum
 error per every eight consecutive quantum registers. \label{Ex:4-reg-code}
\end{example}
\begin{proof}
 Suppose ${\cal E}$ and ${\cal E}'$ be two quantum errors affecting at most
 one quantum register per every eight consecutive ones. By considering $\langle
 \vec{k}'_{\rm encode} | {\cal E}'^{\dag} {\cal E} | \vec{k}_{\rm encode}
 \rangle$, I know that at least six of the following eight equations hold:
 \begin{equation}
  \left\{ \begin{array}{rcl} p_{2i-1} + p_{2i-2} & = & p'_{2i-1} + p'_{2i-2} \\
  p_{2i-1} + p_{2i-2} + q_{2i-2} & = & p'_{2i-1} + p'_{2i-2} + q'_{2i-2} \\
  q_{2i-1} + q_{2i-2} & = & q'_{2i-1} + q'_{2i-2} \\ q_{2i-1} + q_{2i-2} +
  p_{2i-1} & = & q'_{2i-1} + q'_{2i-2} + p'_{2i-1} \\ p_{2i} + p_{2i-1} & = &
  p'_{2i} + p'_{2i-1} \\ p_{2i} + p_{2i-1} + q_{2i-1} & = & p'_{2i} + p'_{2i-1}
  + q'_{2i-1} \\ q_{2i} + q_{2i-1} & = & q'_{2i} + q'_{2i-1} \\ q_{2i} +
  q_{2i-1} + p_{2i} & = & q'_{2i} + q'_{2i-1} + p'_{2i} \end{array} \right.
  \label{E:Systems_2}
 \end{equation}
 for all $i\in\bbbz^+$. Again, I regard $p_i$ and $q_i$ as unknowns; and $p'_i$
 and $q'_i$ as arbitrary but fixed constants. Then, it is straight forward to
 show that choosing {\em any} six equations in Eq.~(\ref{E:Systems_2}) for each
 $i\in\bbbz^+$ would result in a consistent system having a unique solution of
 $p_i = p'_i$ and $q_i = q'_i$ for all $i\in\bbbz^+$. Consequently,
 \begin{eqnarray}
  & & \langle \vec{k}'_{\rm encode} | {\cal E}'^{\dag} {\cal E} |
  \vec{k}_{\rm encode} \rangle \nonumber \\ & = & \sum_{\{ \vec{p},\vec{q} \}}
  \left\{ \prod_{i=1}^{+\infty} \left[ \,\omega_N^{\sum_{j=2i-1}^{2i} p_j (k_j
  + k_{j-2} - k'_j - k'_{j-2}) + q_j (k_j + k_{j-1} + k_{j-2} - k'_j - k'_{j-1}
  - k'_{j-2})} \right. \right. \nonumber \\ & & ~~~~\left. \left. \times\langle
  f_i | {\cal E}'^{\dag}_i | f_i \rangle \,\langle g_i | {\cal E} | g_i \rangle
  \raisebox{1.5em}{~} \!\!\right] \raisebox{1.8em}{~} \!\!\right\}
  \label{E:Systems_3}
 \end{eqnarray}
 for some linearly independent functions $f_i (\vec{p},\vec{q})$ and $g_i (
 \vec{p},\vec{q})$.
 \par
 Now, I consider a basis $\{ h_i (\vec{p},\vec{q}) \}$ for the orthogonal
 complement of the span of $\{ f_i, g_i \}_{i\in\bbbz^+}$. By summing over all
 $h_i$s while keeping $f_i$s and $g_i$s constant in Eq.~(\ref{E:Systems_3}),
 one ends up with the constraints that $k_i = k'_i$ for all $i\in\bbbz^+$.
 Thus,
 \begin{eqnarray}
  & & \langle \vec{k}'_{\rm encode} | {\cal E}'^{\dag} {\cal E} |
  \vec{k}_{\rm encode} \rangle \nonumber \\ & = & \delta_{\vec{k},\vec{k'}}
  \sum_{\{ \vec{p},\vec{q} \}} \left[ \prod_{i=1}^{+\infty} \left( \langle f_i
  (\vec{p},\vec{q}) | {\cal E}'^{\dag} | f_i (\vec{p},\vec{q}) \rangle
  \,\langle g_i (\vec{p},\vec{q}) | {\cal E} | g_i (\vec{p},\vec{q}) \rangle
  \right) \right] ~. \label{E:Systems_4}
 \end{eqnarray}
 Hence, Eq.~(\ref{E:14QCC}) corrects up to one quantum error per every eight
 consecutive quantum registers. \qed
\end{proof}
\par
 From the discussion following Example~\ref{Ex:Perfect_QCC}, the encoding in
 Eq.~(\ref{E:21_QSpin}) can be done efficiently with the help of reversible
 computation \cite{Pebble,SpaceTime,Reversible}.
\section{Outlook}
\label{S:Outlook}
 It is instructive to investigate the coding ability of QCCs as compared to
 that of QBCs. Knill and Laflamme \cite{Cond1} proved that it is impossible to
 construct a four qubit QBC that corrects one general quantum error. Their
 result can be extended to the case when $N > 2$ \cite{5-reg}. Here, with a
 slight modification of Knill and Laflamme's proof, I show that:
\begin{theorem}
 It is not possible to construct a QCC which corrects one general quantum error
 for every four consecutive quantum registers. \label{Thrm:No-4reg}
\end{theorem}
\begin{proof}
 Clearly, the QCC must be of rate 1/4. And with a simple permutation of the
 quantum registers, a general QCC of rate 1/4 can be written as
 \begin{equation}
  |\vec{k}\rangle \longmapsto |\vec{k}_{\rm encode}\rangle \equiv
  \sum_{\{\vec{w},\vec{x},\vec{y},\vec{z}\}} \!a^{(\vec{k})}_{\vec{w},\vec{x},
  \vec{y},\vec{z}} |\vec{w},\vec{x},\vec{y},\vec{z}\rangle ~.
  \label{E:General_Rate1/4}
 \end{equation}
 Without lost of generality, I may assume that quantum errors occurs in
 {\em any} one of the following four set of registers: $|\vec{w}\rangle$,
 $|\vec{x}\rangle$, $|\vec{y}\rangle$ and $|\vec{z}\rangle$.
 \par
 Then, following Knill and Laflamme \cite{Cond1} by considering the action of
 errors in the above four sets of registers, one arrives at $\rho^{(\vec{k})}
 \,\rho^{(\vec{k}')} = 0$ and $\rho^{(\vec{k})} = \rho^{(\vec{k}')}$ for all
 $\vec{k}\neq\vec{k}'$. where
 \begin{equation}
  \rho^{(\vec{k})}_{\vec{w}',\vec{x}';\vec{w},\vec{x}} = \sum_{\{ \vec{y},
  \vec{z} \}} \bar{a}^{(\vec{k})}_{\vec{w}',\vec{x}',\vec{y},\vec{z}}
  \,a^{(\vec{k})}_{\vec{w},\vec{x},\vec{y},\vec{z}} ~. \label{E:Rho_Def}
 \end{equation}
 Hence, the reduced (Hermitian) density matrices $\rho^{(\vec{k})}$ are
 nilpotent for all $\vec{k}$. This is possible only if
 $a^{(\vec{k})}_{\vec{w},\vec{x},\vec{y},\vec{z}} = 0$ for all $\vec{k},\vec{w}
 ,\vec{x},\vec{y},\vec{z}$. This contradicts the assumption that $|
 \vec{k}_{\rm encode}\rangle$ is a QCC. \qed
\end{proof}
\par
 It is, however, unclear if QCC can perform better than QBC in other
 situations. And further investigation along this line is required.
\par\bigskip\noindent
{\em Acknowledgments:} I would like to thank T.~M. Ko for introducing me the
 subject of convolutional codes. I would also like to thank Debbie Leung, H.-K.
 Lo and Eric Rains for their useful discussions. This work is supported by the
 Hong Kong Government RGC grant HKU~7095/97P.

\end{document}